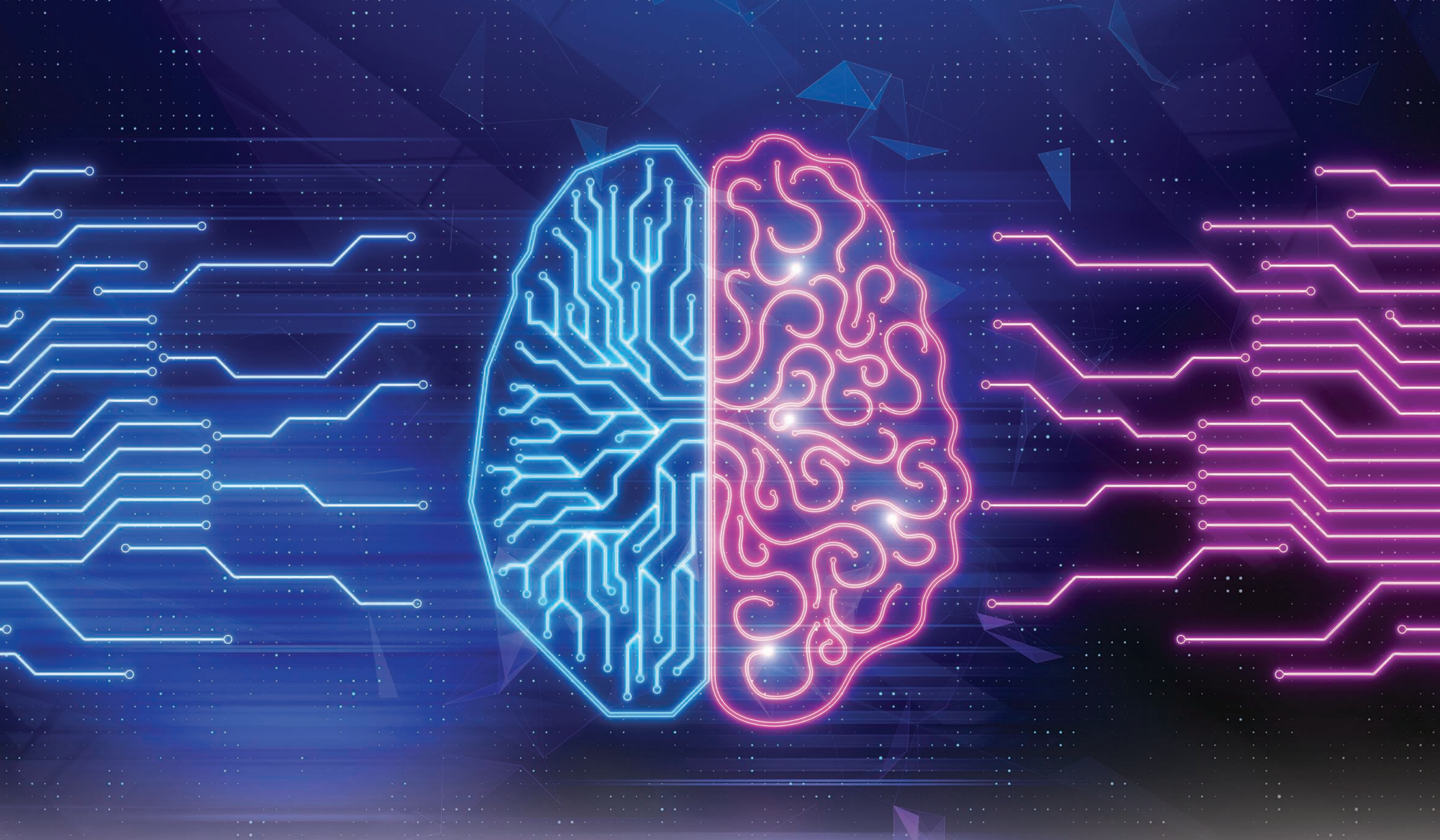

# BeSense: Leveraging WiFi Channel Data and Computational Intelligence for Behavior Analysis

**_Yu Gu and Xiang Zhang_**
*School of Computer & Information, Hefei University of Technology, Hefei, CHINA*

**_Zhi Liu_**
*Dept. of Mathematical & Systems Engineering, Shizuoka University, Shizuoka, JAPAN*

**_Fuji Ren_**
*Dept. of Information Science & Intelligent Systems, University of Tokushima, Tokushima, JAPAN*

***Abstract***—The ever-evolving informatics technology has gradually bounded human and computer in a compact way. Understanding user behavior becomes a key enabler in many fields such as sedentary-related healthcare, human-computer interaction and affective computing. Traditional sensor-based and vision-based user behavior analysis approaches are obtrusive in general, hindering their usage in real-world. Therefore, in this article, we first introduce the WiFi signal as a new source instead of sensor and vision for unobtrusive user behaviors analysis. Then we design BeSense, a contactless behavior analysis system leveraging signal processing and computational intelligence over WiFi channel state information. We prototype BeSense on commodity low-cost WiFi devices and evaluate its performance in real-world environments. Experimental results have verified its effectiveness in recognizing user behaviors.

Corresponding Author: Zhi Liu (E-mail: liu@ieee.org)

## I. Introduction

Nowadays, the ever-changing informatics technology urges people living in modern society to be in a tightly bounded state with a computer. While this coherent state facilitates the accelerating rhythm of urban life and work efficiency, it has gradually tied people to chairs and deprived their exercise time. Its side effects like Sedentary Behavior (SB) pose great threats to people's wellness [1]. Therefore, understanding user behavior, like knowing whether the user is working, gaming or surfing and how long s/he has been doing it, emerges as a key to spot such health risk factors. Moreover, it constitutes a promising enabler to many other fields like Human-Computer Interaction (HCI) and Affective Computing (AC).

In general, there are two types of approaches to understanding user behavior. The traditional vision-based approach leverages video as the main source, where cameras are deployed to record and infer user behavior [2]–[4]. The vision-based approach is effective because of the mature Computer Vision (CV) technology. However, it makes users to concern about their privacy. Also, inherent defects of CV like line-of-sight and illumination constraints further jeopardize its usage in practice. Sensor is another typical source, where wearable sensors are attached to the human body to capture body gestures and deduce the corresponding user behavior [5]–[7]. Sensor has a limited sensing range, and thus multiple sensors are needed at different parts of the human body to ensure complete coverage of user gestures. Unfortunately, such deployment is obtrusive for the user, hindering its practicability in real-world scenarios.

In this article, we introduce the WiFi signal, which is insensible to users, as an alternative source to vision and sensor for perceiving user behavior. The key reason behind is that the human body reflects or absorbs the WiFi signal, and thus changes the WiFi Channel State Information (CSI) [8]–[10]. The inherent research problem is ***how to exploit WiFi CSI that contains rich behavior information to retrieve micro-gestures like keystrokes and mouse movements for understanding the corresponding user behavior?***

Our response to the question is three-fold. Firstly, we explore signal processing to improve the sensing granularity of WiFi CSI. In particular, we build a Fresnel-zone based model to guide the antenna deployment to enhance minor signal changes caused by user's micro-gestures. Then, we design a light-weight segmentation algorithm to extract the micro-gestures from WiFi CSI automatically. The basic idea is to isolate the signal fluctuations caused by the micro-gestures using the stationary state as the benchmark. Lastly, we leverage computational intelligence to recognize these micro-gestures and then understand their corresponding user behavior. We prototype BeSense with low-cost WiFi devices and verify its performance in real environments. Extensive experiments demonstrate that BeSense is very effective in capturing and recognizing user's micro-gestures as well as understanding the corresponding user behavior.

We summarize the main contributions of this article as follows.

- We introduce the WiFi signal as a new source for user behavior analysis. To the best of our knowledge, this is the first work to recognize user behavior using WiFi.
- We leverage signal processing to build a Fresnel-zone based model to enhance insignificant signal changes caused by micro-gestures of users. Then we design a light-weight segmentation algorithm to extract micro gestures from the continuous signal. Lastly, we exploit computational intelligence to recognize these micro-gestures as well as their corresponding user behaviors.
- We prototype BeSense on low-cost WiFi devices and verify its performance in real environments. We also study alternative experimental materials for the human body to accommodate broader real-world scenarios.

The remainder of this paper is organized as follows: in the next section, we provide an overview of the related works. We introduce the system design in section III, and evaluate the experimental results in section IV. We discuss the effect of different objects on the received CSI signal in section V. Finally, we conclude our work and discuss some open issues in section VI.

## II. Related Works

WiFi-based behavior sensing technology has many advantages over traditional behavior sensing technology (e.g., vision-based sensing technology, infrared-based sensing technology and dedicated sensor based sensing technology) in terms of non-line-of-sight, passive sensing (no need to carry sensors), low cost, easy deployment, no restrictions on lighting conditions, and strong scalability. A large number of applications have emerged based on existing WiFi signals. From daily behavioral awareness [11], [12] and gesture recognition [13], [14] to identity authentication [15], [16] and from individual physiological indicators [17], [18] to group perception [19], [20] and fall detection [21], [22], behavior sensing technology based on WiFi is showing unprecedented potential for application, achieving not only the interaction between machines and machines but also the natural interaction between humans and machines.

In the early days, WiFi-based behavior sensing mainly uses Received Signal Strength (RSS). Sigg *et al.* [23] use a software radio to transmit RF signals and determine human motion based on changes in RSS. Abdelnasser *et al.* leverage RSS to identify 7 different gestures [13] and respiratory detection [24]. We also built a similar RSS-based system PAWS to handle whole-body activities [8]. Since RSS is coarse-grained and CSI can yield more detailed information, recent research mainly uses CSI for behavior sensing. WiFall [25] uses CSI to implement the fall detection system. Zeng et al. [19] leverage CSI to recognize five different customer behavior states. We also built a CSI-based system MoSense to pinpoint the motions in a real-time manner [9].

These schemes basically rely on high-level features like location, velocity and direction of a motion for behavior recognition. In particular, WiSee [26] and WiDance [27] use the Doppler shift to extract the direction of motion and then use this direction information to classify different motions. WiDir [28] uses Fresnel Zone theory to extract direction and distance information to identify the direction of walking. CRAM [12] uses time-frequency analysis and Discrete Wavelet Transformation (DWT) to extract velocity information and Hidden Markov Model (HMM) to achieve behavior recognition. Using high-level features for behavioral recognition is more reasonable than using statistical features.

**In particular, we build a Fresnel-zone based model to guide the antenna deployment to enhance minor signal changes caused by user's micro-gestures.**

Since human behavior is complex and fuzzy in nature, recently there is a trend to explore computational intelligence, which is a specialized paradigm to deal with this kind of problems [29]–[31]. These methods extract features or directly input waveform data into the deep network model and train the network model for motion recognition. For example, Li *et al.* adopt a multi-layer Convolutional Neural Network (CNN) for learning human activities using WiFi CSI from multiple Access Points (APs) [29]. This deep learning-based recognition has higher accuracy than traditional solutions but requires substantial training data and training time. They are also unable to deal with micro-gestures since the corresponding changes in WiFi CSI are insignificant or even trivial sometimes.

To this end, we present BeSense, which leverages signal processing to handle the micro-gestures and computational intelligence for behavior analysis. It extends our previous work [32] with non-trivial improvement. In particular, we have improved the signal segmentation algorithm, which results in better performance than the previous method based on variance thresholds. In contrast to the previous gesture-frequency-based behavior recognition method, we adopt an HMM-based recognition method for behavior recognition, which can provide better performances. We also conduct a series of experiments to find alternatives with similar influences on the channel response to simulate human movements.

## III. BeSense: System Design and Analysis

The WiFi signal possesses several attempting merits and thus constitutes a promising source for user behavior analysis. However, before we can fully enjoy its benefits, we need to handle one major challenge associated with it: ***how to exploit WiFi CSI that contains rich behavior information to retrieve micro-gestures for understanding the corresponding user behavior?***

To this end, we leverage signal processing and computational intelligence to design BeSense, which consists of three layers as shown in Fig. 1:

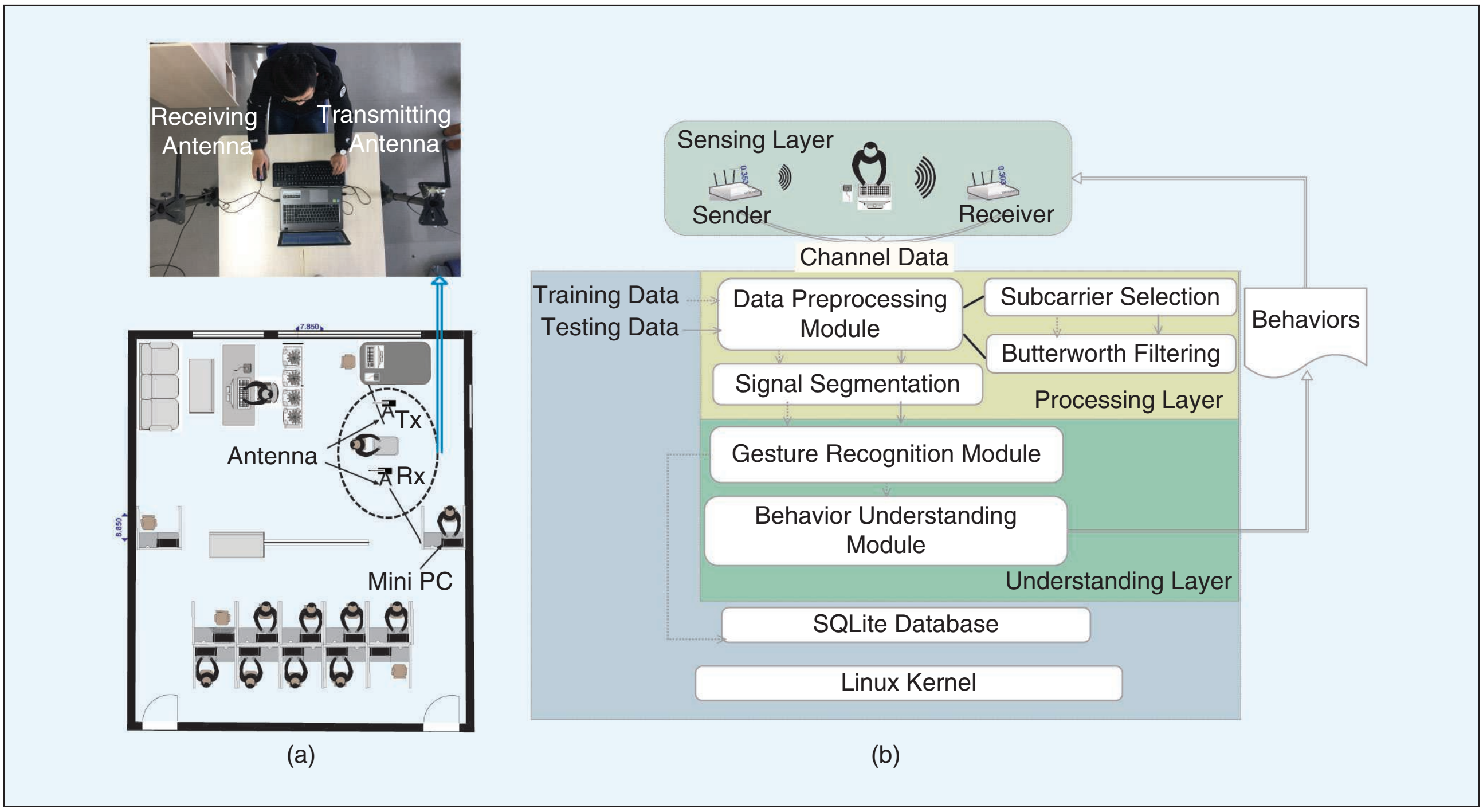

FIGURE 1 Hardware setup and system architecture of BeSense: The left is hardware placement diagram; the upper-left is the experimental reality; the right is the system architecture diagram.

**We design a light-weight segmentation algorithm to accurately extract these short signal segments corresponding to the micro-gestures by using the stationary state as the benchmark.**

- **Sensing layer.** A micro-gesture is small in range. Therefore, the corresponding change in the signal by body reflecting or absorbing is also small, or even trivial sometimes. To deal with this issue, we explore signal processing and build a Fresnel-zone based model to guide the antenna deployment to enhance such signal changes.
- **Processing layer.** A micro-gesture is short in time. After signal enhancement, we design a light-weight segmentation algorithm to accurately extract these short signal segments corresponding to the micro-gestures by using the stationary state as the benchmark.
- **Understanding layer.** User behavior consists of a serial of micro-gestures, which require to be recognized. We leverage computational intelligence for recognizing micro-gestures and the corresponding user behavior.

### A. Sensing Layer

A typical user micro-gesture like keystroke and mouse movement ranges from 2 to 5 cm, which is very difficult to be captured on WiFi CSI. Traditional CSI-based motion detection systems like MoSense [9] are incapable of such micro-gestures.

In our previous work EmoSense [33], we show that the antenna deployment is important for capturing body gestures, especially for micro-gestures. Therefore, in the sensing layer, we explore signal processing and build a Fresnel zone-based model to guide the antenna deployment.

Fig. 2 demonstrates what a Fresnel zone is and how it is used to improve the sensitivity of the WiFi signal. In Fig. 2(a), $P_1$ and $P_2$ represent the transmitting and receiving antenna, respectively. Fresnel zones are a series of concentric ellipsoid, which can be constructed via the following equation for a given signal wavelength $\lambda$,

$$|Tx, Q_n| + |Q_n, Rx| - |Tx, Rx| = n\lambda/2, \qquad (1)$$

where $Q_n$ is a point at the boundary of the $n$th Fresnel zone, and $|x, y|$ calculates the physical distance between position $x$ and $y$.

Channel Frequency Response (CFR) can be expressed simply as the superposition of dynamic path CFR and static CFR, and it can be represented as:

$$H(f, t) = H_s(f, t) + H_d(f, t). \qquad (2)$$

The dynamic CFR can be written as:

$$H_d(f, t) = \sum_{k \in D} h_k(f, t)\, e^{-j2\pi f \tau_k(t)}, \qquad (3)$$

where $f$ and $\tau_k(t)$ represent the carrier frequency and the propagation delay on the $k$th path, respectively. $D$ is the set of dynamic paths. When a WiFi signal is reflected by a subject and the subject moves a small distance, it will lead to changes in the phase of the WiFi signal on the corresponding path. If the subject moves $d(t)$, since wireless signals travel at the

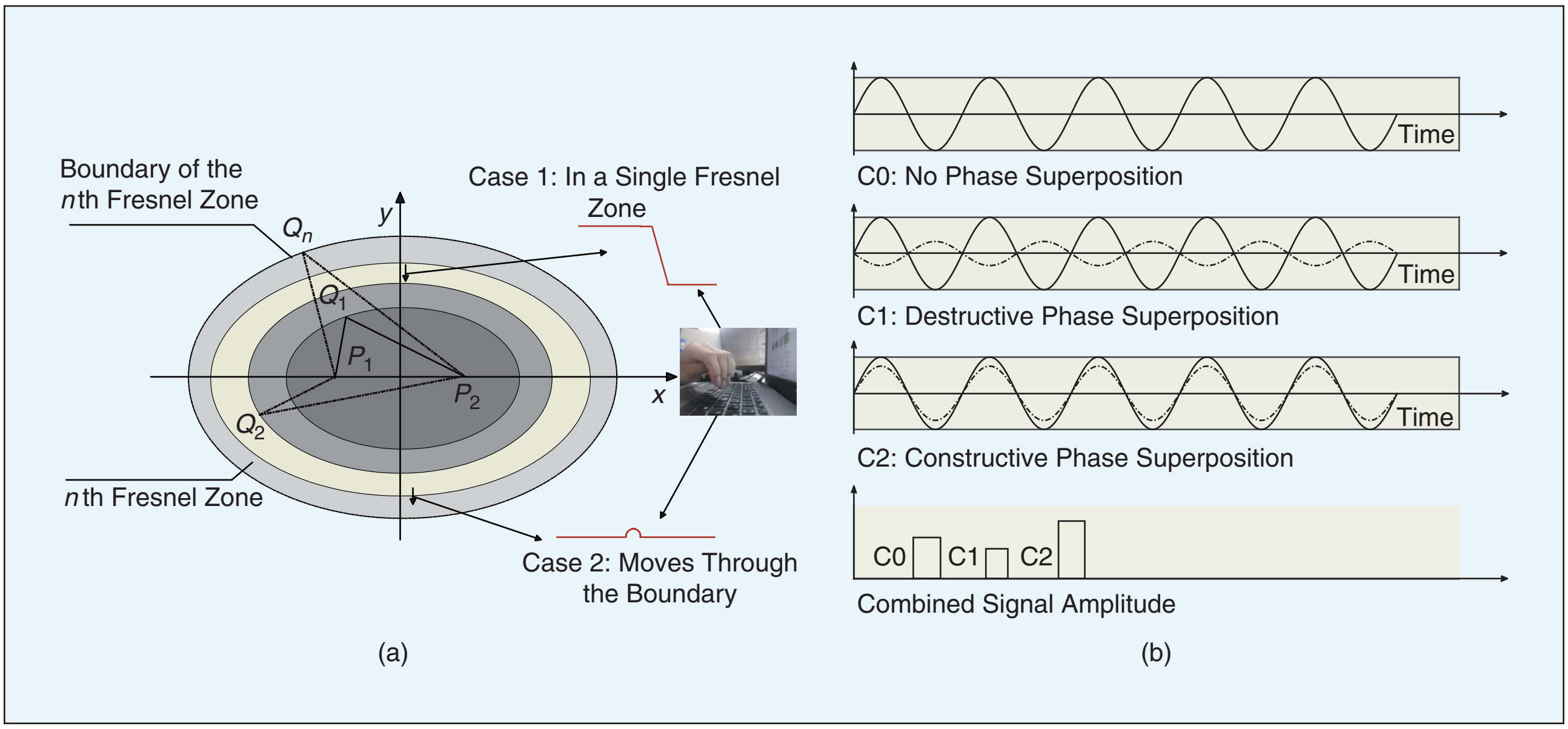

**FIGURE 2** (a) Fresnel Zone; (b) Signal superposition.

speed of light, denoted as $c$, $\tau_k(t)$ can be represented as $d(t)/c$. Let $\lambda$ represent the wavelength, where $\lambda = f/c$. Thus, the phase shift can be written as $e^{-j2\pi d(t)/\lambda}$. Therefore, when a subject appears at the boundary of the even/odd Fresnel zone, the dynamic path phase shift $\Delta p$ is equal to $\pi$ and $2\pi$, respectively. As a result, the combined signal amplitude should be **degraded** in the even zones and **enhanced** in the odd zones, as shown in Fig. 2(b).

A gesture consists of directed motions. For example, a keystroke consists of two directed motions, i.e., finger moving down and up. We divide the directed motions into two cases according to whether the motion crosses the Fresnel zone boundary:

**Case 1: It happens within a single Fresnel zone.** In this case, the superposition signal should be either monotonically increasing or decreasing.

**Case 2: It involves two Fresnel zones.** In this case, the superposition signal does not show such monotonicity.

Fig. 2(a) presents a case study, where one keystroke is captured under both cases. When it only happens in a single zone, i.e., the 3rd zone, the amplitude of the superposition signal is monotonically decreasing. However, when it crosses two zones, i.e., the 3rd and 4th zones, the amplitude of the superposition signal increases first and then decreases.

Therefore, it is feasible to keep the micro-gestures within some single zones by an appropriate antenna deployment. Particularly, we select the 9th, 10th and 11th zones since their thickness is around 4 cm, close to our target micro-gestures like keystrokes and mouse movements.

### *B. Processing Layer*

The processing layer processes raw CSI data to point out signal segments corresponding to micro-gestures. More specifically, it first selects subcarriers with finer granularity, and then denoise them using Butterworth filter. Lastly, a light-weight segment algorithm is designed for finding out the signal segments of micro-gestures.

**Subcarrier Selection.** According to [34], different subcarriers have different sensitivities to human motions. This is because different subcarriers have different central frequencies and wavelengths. Combining the effect of multipath/shadowing with different frequencies, CSI measurements for one motion at different subcarriers have different channel responses. Therefore, it is essential to choose proper subcarriers that can better capture the designated gestures.

According to our empirical studies, we find that the channel response for the same motion differs significantly for subcarriers with large sequence number difference. Fig. 3 shows such an example in which we visualize one experiment involving both typing and mouse moving that lasts for approximately 10 seconds. It can be clearly seen from the heatmap that the first 10 subcarriers are much more capable of preserving gestures than the last 10. To further illustrate the phenomenon, we select the 4*th* and 30*th* subcarriers for comparison, where we can observe that the channel response on the latter is too weak to seize the attenuation caused by gestures. In general, the larger the variance of the subcarrier waveform, the more sensitive it is to the action. Therefore, we select the subcarrier, which has the largest variance in our experiments.

**Butterworth Filter.** The raw channel data selected may contain abnormal samples caused by background noise or hardware glitches and thus should be filtered. In the preprocessing module, we choose the Butterworth filter to denoise the data. Under our experimental conditions, the speed at which we type or move the mouse is usually between 2 and 60 cm/s. One second of the movement may pass through approximately 15 Fresnel zones, which will go through 15 peaks or nadirs; therefore, the cutoff frequency of the filter is preferably set to 7.5 Hz.

**Signal Segmentation.** The major challenge of segmentation lies in the short duration of each designated gesture. For example, a keystroke usually only lasts for approximately 0.7 s. To ensure real-time recognition, the segmentation algorithm should be fine-grained, adaptive and light-weight. Here, we leverage the variance of the channel data, which differs significantly between the motion and stationary state, and design an automatic segmentation algorithm with computational intelligence, as shown in Alg. 1. The key steps of this algorithm are as follows:

- **Variance calculating.** Calculate the variance of the waveform with the window size 1/20 s and step size $1/fs$ s, where $fs$ is the rate of packet sending. This calculation will convert the original waveform into a waveform of variance. When the original waveform is relatively calm, the amplitude of variance waveform is close to 0, and when the original waveform fluctuates greatly, the amplitude of variance waveform is much larger than 0. Compared with the original waveform, this variance waveform retains the fluctuation characteristics and is not affected by the change of the amplitude. We use $nor1$ to represent this variance of the waveform, as shown in Fig. 4(b);

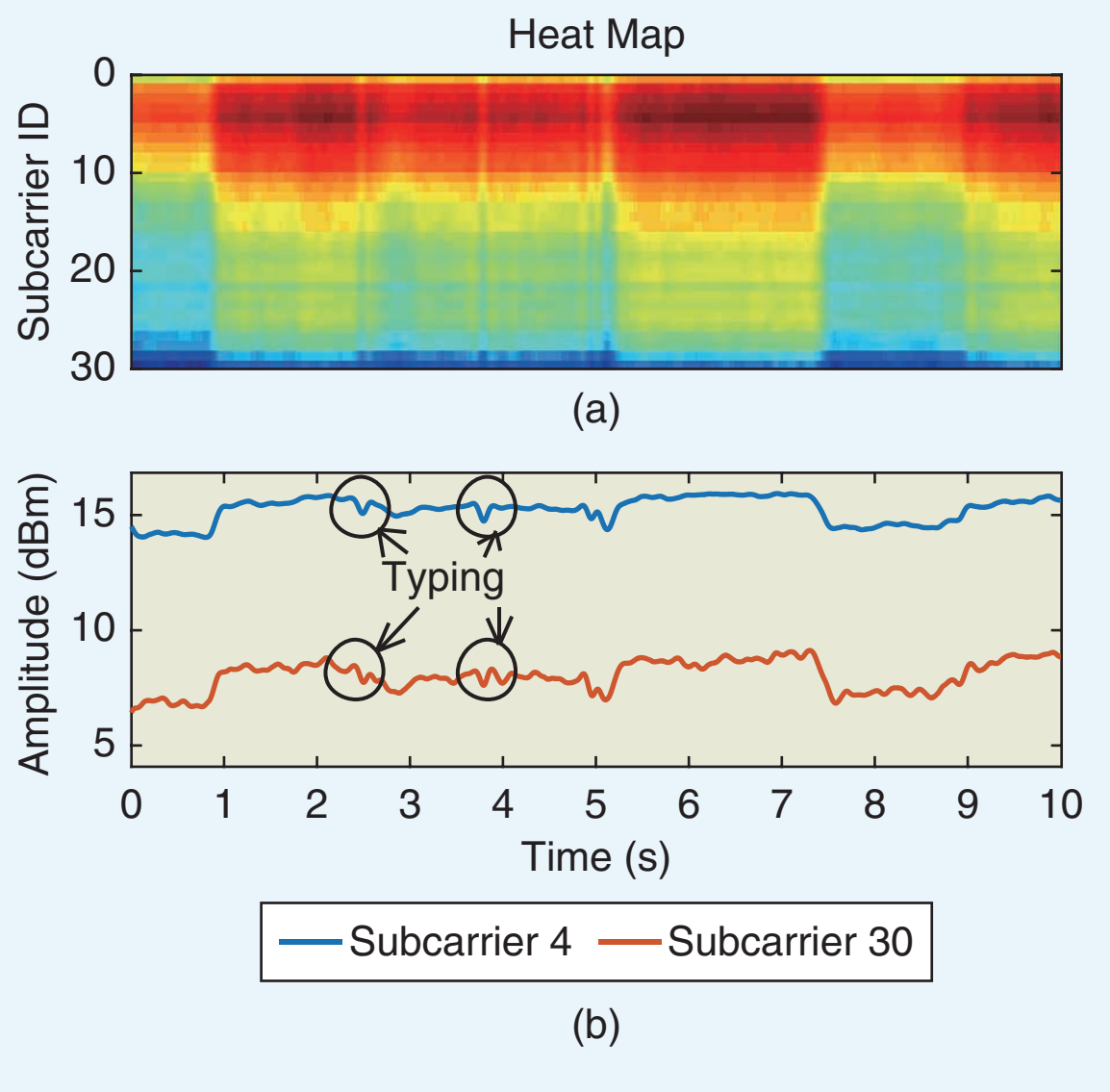

**FIGURE 3** Subcarrier difference: In the above figure, the channel response of the first ten subcarriers is more obvious than others; In the below figure, for the same typing gesture, subcarrier 30 has a weaker channel response than subcarrier 4.

**Algorithm 1 Automatic segmentation algorithm.**

```
Input: CSI_f
Output: Gesture Waveform
begin
  windows = 1/20 s;
  step = 1/fs s;
  nor1 = variance(CSI_f, windows, step);
  nor2 = 100 * variance(sum(nor1), windows, step);
  while t < length(nor2) do
    set = ∅;
    for se = 0.1:0.1:5 do
      for i = t : length(nor2) do
        se = se + nor1(i) − nor2(i);
        if se < 0 then
          set = [set; i];
        end
      end
    end
    for j = 1: length(set) do
      if max(set(j : j + 5) − min(set(j : j + 5)) < 10/fs s
      then
        start − point = set(j);
        e = nor2(set(j));
        for x = set(j) : length(nor2); do
          if nor2(x) <= e then
            end − point = x;
          end
        end
        t = end − point; end
      end
    end
  forall the w in waveform segmented do
    if max(w) − min(w) < threshold then
      remove it;
    end
end
```

- **Data smoothing.** For $nor1$, small fluctuations exist in the stationary state due to noise interference. To denoise these fluctuations, we first perform a sum operation on $nor1$ with the same window size 1/20 s and step size $1/fs$ s. Then similar to step 1 (i.e., variance calculating), we calculate the variance of the waveform obtained after the sum operation. After such operation, we multiply the newly generated waveform by 100 and then obtain $nor2$, as shown in Fig. 4(c);
- **Star-point marking.** We compare $nor1$ to $nor2$, as shown in Fig. 4(d), and observe that the denoising is effective and that the difference between $nor1$ and $nor2$ of no gesture states is very small. But when the gesture occurs, the difference between $nor1$ and $nor2$ is very large, as shown in Fig. 4(d), and we find the start-point of gesture based on this character. Specifically, set an initial value $se$ and set $sp$ as the start point of the motion detection. If $se > 0$, then $se = se - (nor2(sp) - nor1(sp))$, $sp = sp + 1$; otherwise, record the $sp$ as a segment point. $se$ increases from 0.1 to 5 with a step of 0.1 in our experiments; thus, we can obtain 50 segment points. Since when the gesture occurs, the value difference between $nor1$ and $nor2$ is very large. Thus, at the start-point of gesture, increasing $se$ will not make the segment point move back too much, as shown in Fig. 4(d). We select one particular segment point as the start-point; the difference between this point and its five following consecutive segment points is smaller than a threshold ($10/fs$ s in the experiments).
- **End-point marking.** Given that the value of the end-point of each motion is almost the same as its start-point in terms of $nor2$, we find a point $en$ after $spt$ ($spt$ represents the start-point of motion in step 3 (i.e., start-point marking)). If $nor2(en) <= nor2(spt)$, set $en$ as the endpoint of this motion. Go to step 3.
- **Data validating.** We omit the segmentation results with amplitude differences less than a certain value. The results obtained by using our segment algorithm for the original waveform in Fig. 4(a) are shown in Fig. 5. There are 17 subfigures, and each subfigure is a waveform of a typing; the 17 subfigures are for the 17 gestures of Fig. 4(a).

### C. Understanding Layer

**Micro-gesture Recognition.** We use a traditional classifier (such as SVM, KNN and Random Forest) to determine whether the micro-gesture is typing or mouse moving. The feature selection plays a central role, and the features used by the classifier are as follows.

The directions of the two types of micro-gestures are basically orthogonal (one horizontal to the desktop and one perpendicular to the desktop). If we type or move at almost the same speed; however, the signal propagation path length change speed of these two gestures is very different. The intensity of the waveform fluctuations can reflect the speed at which the signal propagation path changes; thus, we use the variance of gesture waveform as the first feature of the classifier. The typing gesture is symmetrical (press the keyboard first and then release the keyboard), which is consistent with the CSI waveform; however, the gesture of moving the mouse is usually asymmetric. We divide the waveform into two parts, calculate the slope of the line from the maximum point to the minimum point of the two parts, and use the largest absolute values of the slope ratio as the second feature, which can characterize the symmetry of the waveform. The duration of different actions will be different; therefore, we use the gesture duration as the third feature. Then we map the segments to the corresponding gestures through the SVM, KNN and Random Forest classifiers to recognize the different gestures.

**We leverage the variance of the channel data, which differs significantly between the motion and stationary state.**

**Behavior Recognition.** Behavior is a composite of gestures. In our case, all the three behaviors (i.e., surfing the internet, gaming and working) consist of two basic features, i.e., typing and mouse moving. However, the ratio of each gesture contained in each behavior is different, as shown in Table I, and they behave differently at different times. To this end, we use a method based on the HMM to recognize behaviors.

The HMM consists of three parts, namely, the state transition probability matrix, emission probability matrix and initial state probability distribution. The probability of generating an observed state in a hidden state at a certain moment is called the emission probability. Here, the actions of typing and moving a mouse are used as the hidden state, and the recognition result of the motion waveform processed by the classifier is used as the observed state. Therefore, we use the accuracy and error rate of the classifier to classify two types of actions as the emission matrix. We count some of the action sequences of using a computer and obtain the probability of using the mouse or keyboard first; then, we take this value as the initial state probability distribution. The Baum-Welch algorithm is used to generate the state transition probability matrix.

We use the above method to establish the HMM for three behaviors including surfing the internet, gaming and working. For other behavior that needs to be identified, we also use the above method to build the HMM, compare the new model with the three existing behavioral models, and select the one with the greatest similarity as the discriminant result.

## IV. Performance Evaluation

This section systematically evaluates the performance of BeSense via real-world experiments.

### *A. Hardware Setup*

We prototype BeSense on commodity WiFi devices. As shown in Fig. 1, we use two identical mini PCs for sending and receiving via Intel 5300 Network Interface Card (NIC) mounted on them. The mini PC is mounted with 2.16 GHz Intel Celeron N2830 processor with 2 GB RAM running on Ubuntu OS version 12.04. We only use one pair of antennas. The spacing between the transmitter and receiver

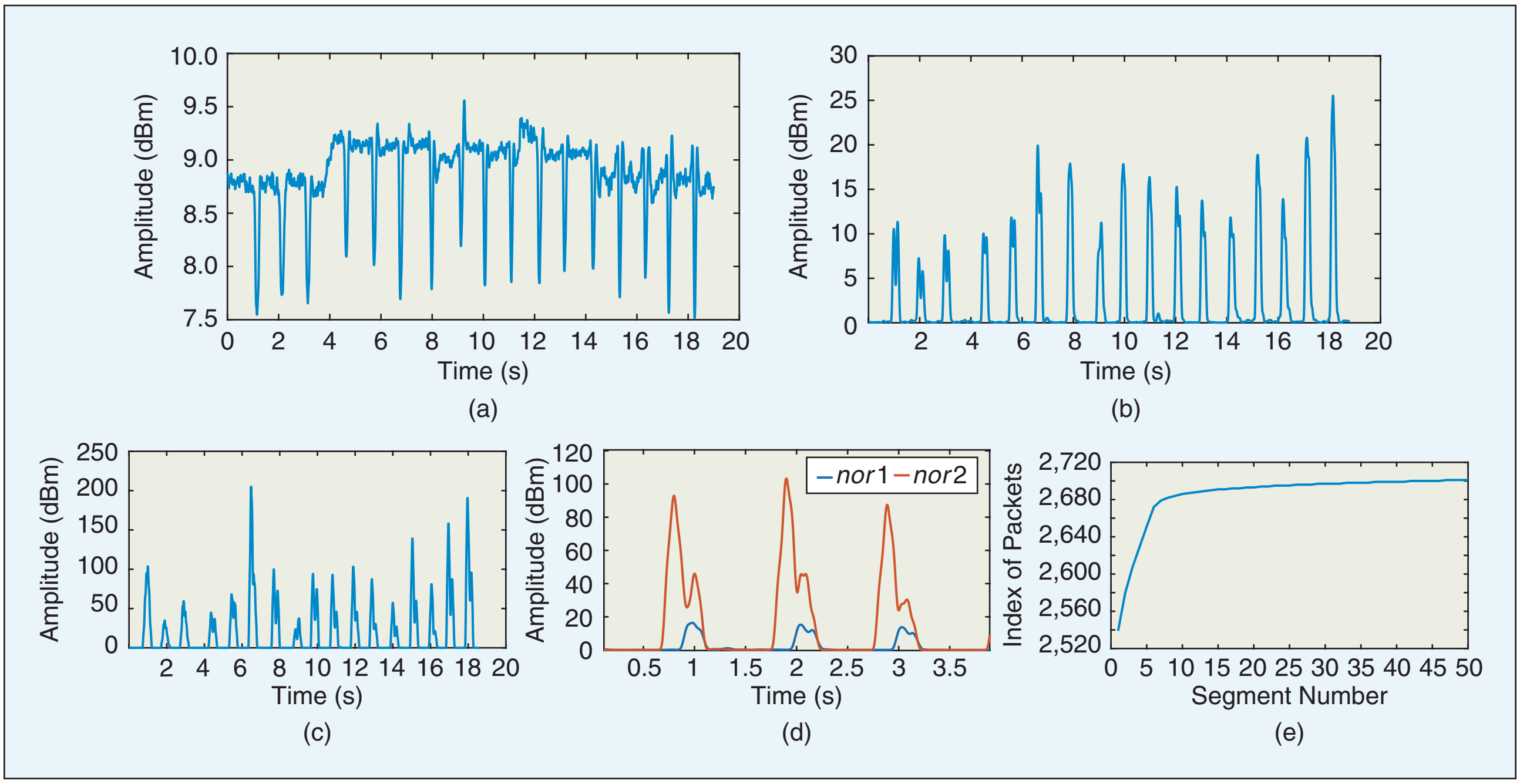

**FIGURE 4** (a) Original CSI waveform of 17 consecutive typing gestures; (b) *nor*1: variance of the original waveform; (c) *nor*2: data smoothing on *nor*1; (d) Comparison of *nor*1 and *nor*2; (e) Start-point determination for 3rd gesture in the original waveform. Since the result tends to be stable after the 7th segment, set the 7th segment point as start-point.

**CSI-based systems are resilient to human activities outside of the area of interest.**

antenna is 1 meter, which is horizontally level above the floor, and our keyboard and mouse are placed 55–70 cm directly below the antenna. The sampling rate is set to 1000 packets per second.

For BeSense, the radius of the Fresnel zones varies between 3.72 cm and 4.14 cm, which is similar to the magnitude of our keystroke (2 cm), and we can obtain better channel data. As shown in our previous work [33], CSI-based systems are resilient to human activities outside of the area of interest. Since BeSense only focuses on a small area of interest, i.e., desktop, we allow the presence of other people in the experimental rooms.

### *B. Experimental Environment and Data Collection*

In our experiments, training and testing datasets were collected from the office environment, as shown in Fig. 1. We used 10 university students (6 females) who volunteered for the experiments. Their ages range from 21 to 26. Since our system design goal is to identify three behaviors (surfing, gaming and working) when people use their computers, each of our testers conducted 8 experiments (three behaviors each performed twice; typing and mouse moving experiment once and each of the latter two experiments contains 10 gestures). Additionally, 5 of the volunteers conducted 15 additional experiments (three kinds of behaviors each performed once) to establish the HMM of each behavior. Finally, we collected 95 data files.

We evaluate this proposed scheme in terms of the accuracy of distinguishing typing/mouse moving movements and the accuracy of distinguishing the following three behaviors: surfing, gaming and working.

**TABLE I** Relative use frequency of keyboard and mouse.

| | KEYBOARD | MOUSE |
|---|---|---|
| STATIC | LOW FREQUENCY | LOW FREQUENCY |
| WEB | LOW FREQUENCY | MEDIUM FREQUENCY |
| WORK | HIGH FREQUENCY | MEDIUM FREQUENCY |
| GAME | HIGH FREQUENCY | HIGH FREQUENCY |

### *C. Typing/Mouse Moving Recognition*

We use 5 classifiers to evaluate our typing/mouse moving movements classification accuracy, which are SVM, KNN, Naive Bayes, Random Forest and discriminant analysis classifier. We use a segmentation algorithm to segment the waveforms, obtain the waveform containing the gestures, extract features, and use a classifier to classify them. In order to obtain accurate results, we divide the gestures of each person into ten parts (each part contains typing one time and mouse moving one time), and then we use 10-fold cross-validation in all experiments to obtain the final result.

For identifying the typing and mouse moving gestures in our experiments, the accuracy of the SVM, KNN, Naive

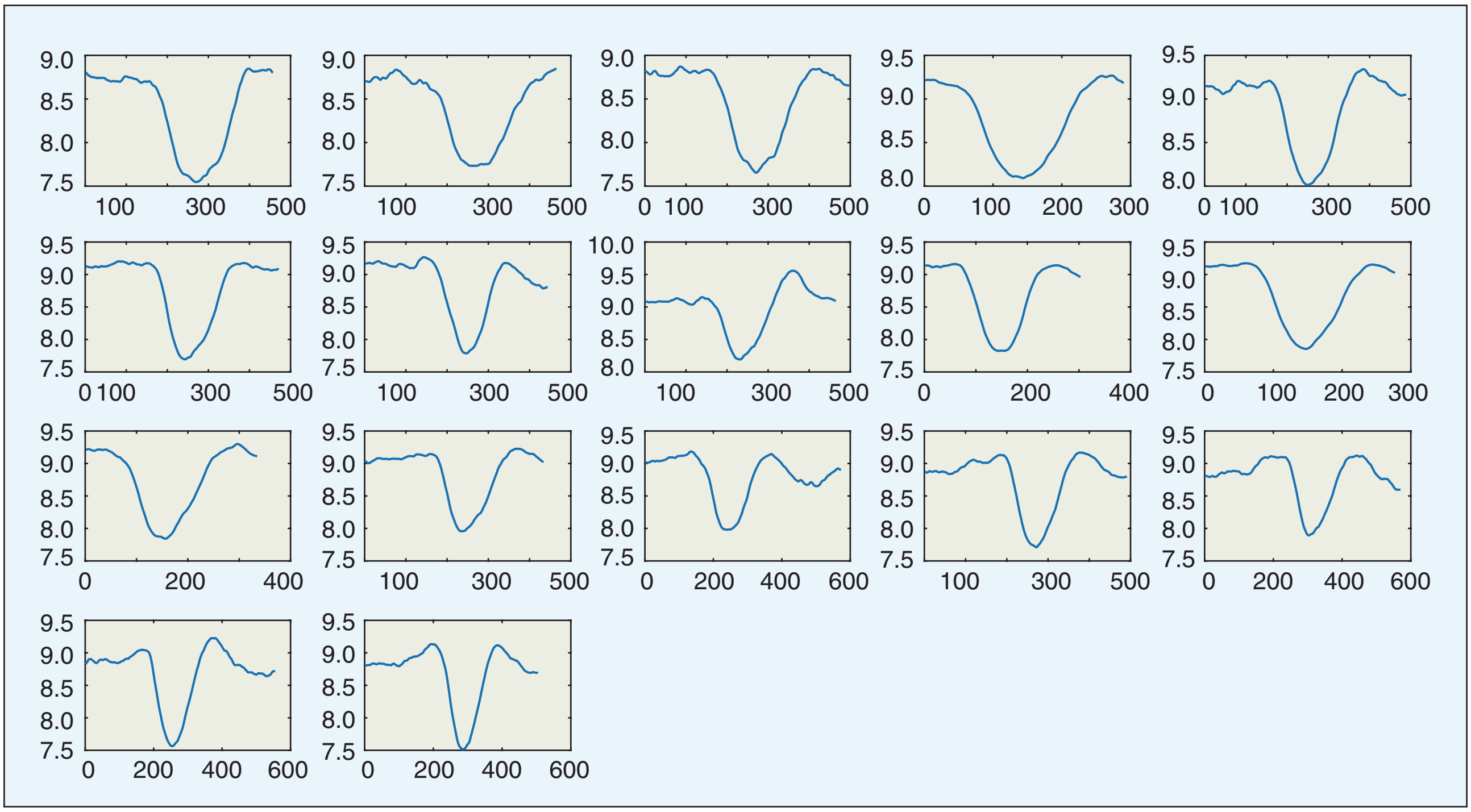

**FIGURE 5** Segment result: 17 subgraphs correspond to the 17 gestures in Figure 4(a), x-axis: number of packets; y-axis: amplitude (dBm).

Bayes, Random Forest and discriminant analysis classifier are 96.5%, 96.5%, 96%, 98%, and 79%, respectively; therefore, SVM, KNN and Random Forest are used in real-world experiments due to their higher accuracy.

**The typical factors influencing the channel responses include material, size and thickness.**

Fig. 6(a) shows the accuracy of distinguishing two gestures using SVM, KNN and Random Forest, where the x-axis represents the users (volunteers) and the y-axis represents the accuracy. Due to the fact that movement habits are different for each person, the accuracy is also different. From the results, we can observe that SVM has high accuracy when classifying mouse moving and that Random Forest performs better in classifying typing gestures, reaching an accuracy of 100%.

### *D. Real-world Evaluation of Behavior Analysis*

In this section, we use all typing and mouse moving experiment data collected from the 10 volunteers as a training set to classify gestures. Then, we train the HMM with the behavior data that needs to be identified, compare the obtained HMM with the previously trained HMM of the three behaviors, and select the most similar data as the recognition result. The final results show that the three behaviors' recognition accuracy values of 93.3% (SVM), 80% (KNN), and 68.3% (Random Forest), respectively. We use the confusion matrix shown in Table II.

Fig. 7(b) shows the recognition accuracy of three behaviors, where the x-axis represents the users and the y-axis represents accuracy. We can find that the recognition accuracy of user 2 is the lowest, only 66.7%; however, the recognition accuracy of the mouse moving/typing of user 2 is good. Moreover, the overall recognition accuracy is not good. This is because the mouse and keyboard are used alternatively and affect each other. When stopping after an action, the waveform will take some time to completely calm down. If the interval between two gestures is not long enough, the previous one will affect the waveform of the following gestures. Both Table II and Fig. 6(a) show that we have the highest recognition rate

**TABLE II** Confusion matrix of three behaviors using SVM.

| | SURFING | WORKING | GAMING |
|---|---|---|---|
| SURFING | 100% | 0% | 0% |
| WORKING | 5% | 95% | 0% |
| GAMING | 15% | 0% | 85% |
| | | AVG. | 93.3% |

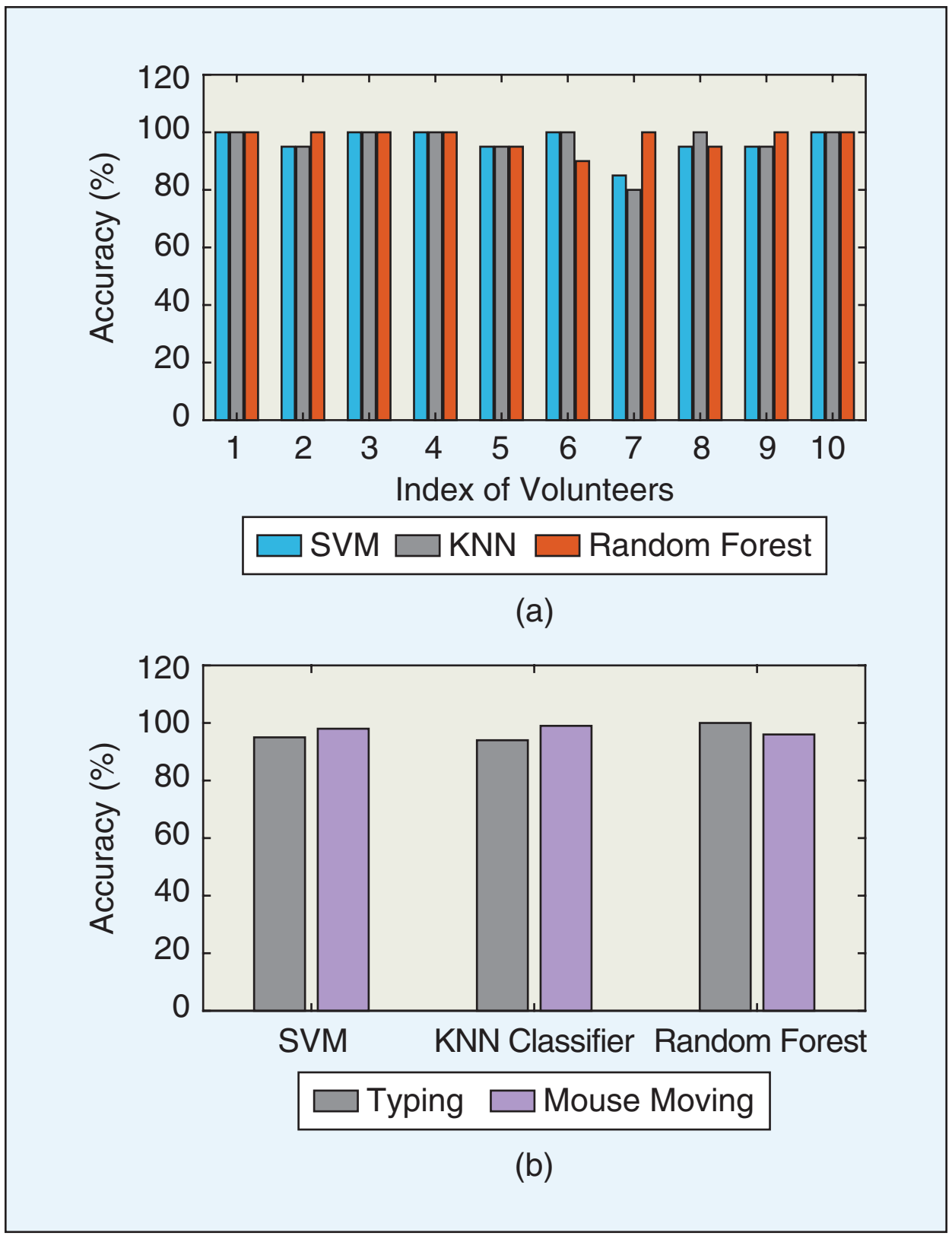

**FIGURE 6** (a) Classification accuracy of different classifiers; (b) Motion classification accuracy.

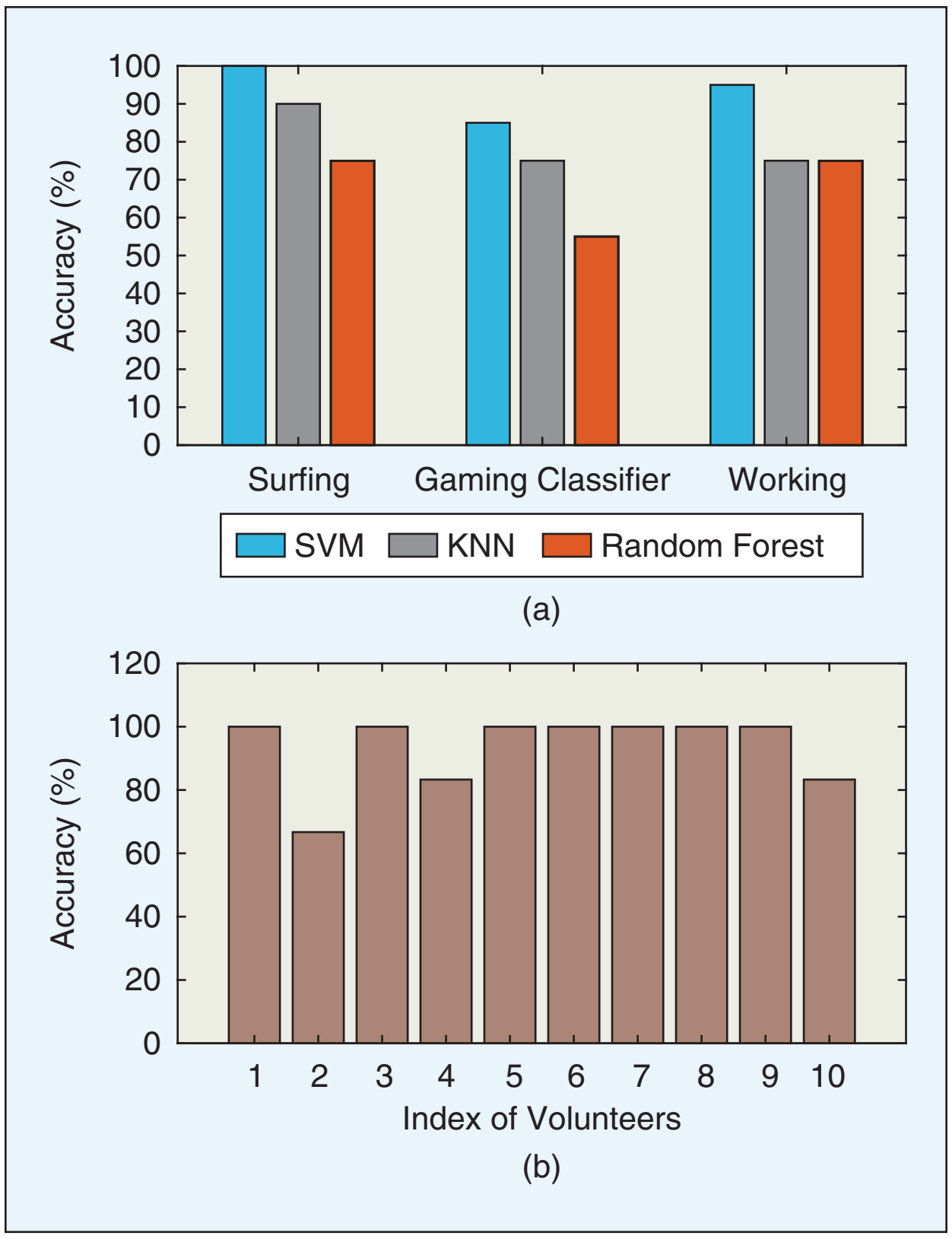

**FIGURE 7** Classification accuracy of (a) different classifiers and (b) different volunteers.

**We envision WiFi playing a vital role in behavior recognition with its ability to offer satisfactory recognition results.**

for surfing and the lowest recognition rate for gaming because the two gestures switch more frequently when playing the game, which affects the classification accuracy of gestures and ultimately affects the accuracy of behavior recognition.

## V. Analysis of Factors Influencing the Micro-Motion Experiment

BeSense currently only focuses on a few most commonly-seen behaviors that can be safely conducted by humans. But its segmentation and classification modules are pervasive and can handle any continues or even composite behaviors. The key problem is that some critical events like fall or seizure could be harmful to the participants. To this end, we present some alternatives to simulate human bodies via empirical studies.

The typical factors influencing the channel responses include material, size and thickness. In order to study the influences of these factors on the channel responses, a series of experiments are designed. In particular, we make square plates with a thickness of 1.5 mm using three materials, i.e., plastic, board and cardboard. The length of each side is 2 cm, 3 cm, …, 12 cm, respectively. In each experiment, the square plate is placed just 50 cm below the middle of two antennas, and it is dragged using a rope from a remote location with a speed of approximately 8 cm/s. The moving direction is the same as the direction between the transmitting and receiving antennas, and the moving range is 1.5 cm, which is similar to that of the mouse movement. All experiments are repeated 20 times.

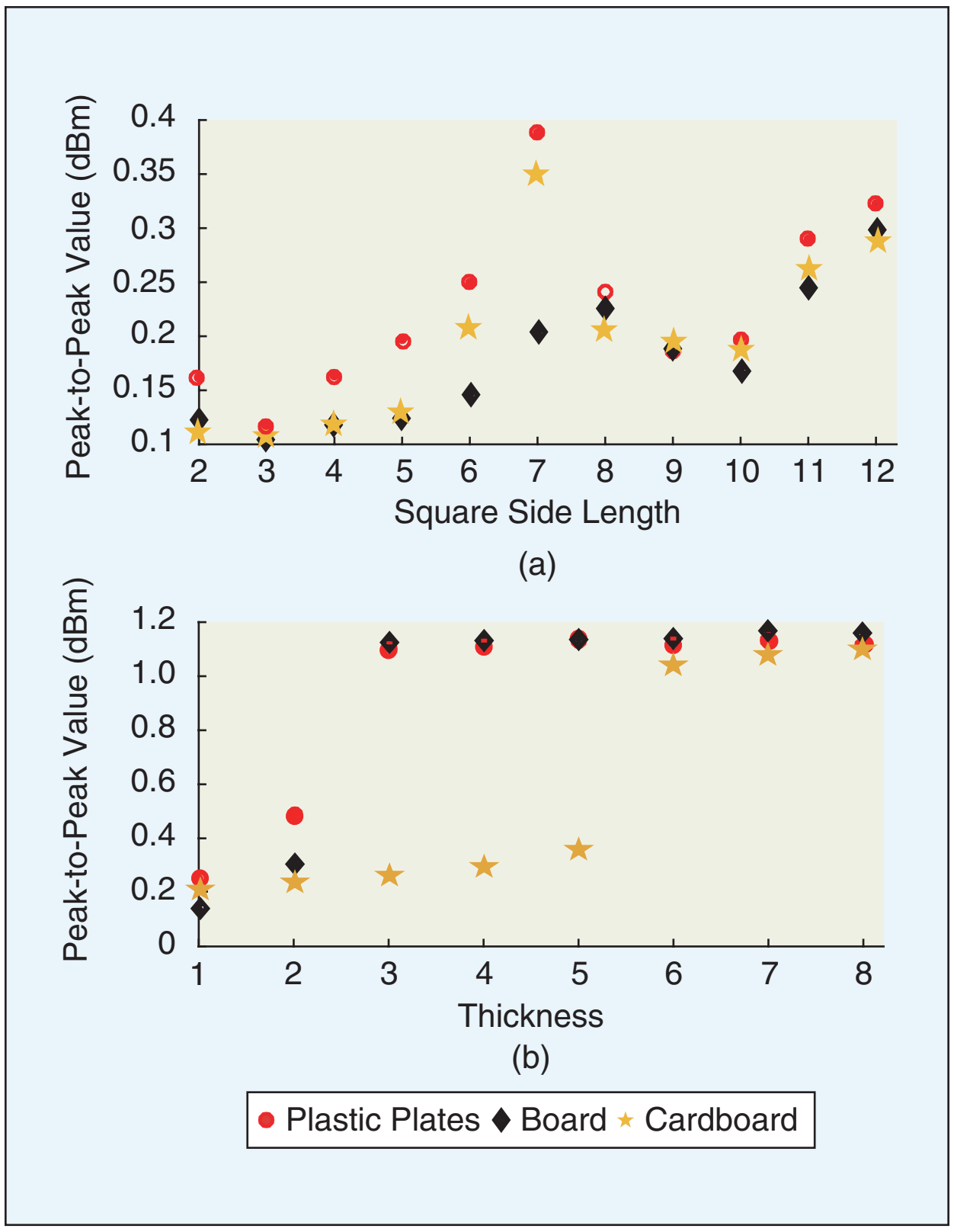

**FIGURE 8** (a) Relationship between object area and channel response; (b) Relationship between object thickness and channel response.

In the first experiments, the 33 square plates are used to study the relationship between object size and channel response. Fig. 8(a) shows the average peak-to-peak values of the obtained waveform, and we can observe that the peak-to-peak values of the received signal do not keep increasing as the size increases. Instead, the peak-to-peak value of the received signal first increases and then decreases before increasing again. This observation could be explained by the theory of the Fresnel zone. As we know, when the size of the moving object increases, the area of the reflected signal increases. In particular, in our micro-motion experiments, the object will cross one Fresnel zone into the adjacent Fresnel zone as the size of the object increases. In the end, the object will have a part in the signal-enhanced Fresnel zone and the rest in the signal-weakened Fresnel zone; thus, the signals will cancel each other out at the receiving end. This also indicates that the object size is not the dominating factor, and the same object with slow motion can also influence the channel responses significantly when the experiment is set up properly.

In the second experiments, 24 square plates with 8 different thicknesses in three materials are exploited to study the relationship between the thickness and channel response; the side length of these plates is 6 cm. Fig. 8(b) shows the average peak-to-peak value of the received signal. From the figure, we can find that the receive signal's peak-to-peak value will keep increasing as the thickness of the object increases. Moreover, increasing the thickness could significantly increase the channel response at a certain thickness. However, the channel response tends to be stable after reaching a certain thickness. According to the experimental results, the relationship between the thickness and channel response could be determined by the penetration, reflection and diffraction of the signals. When the thickness of the object increases to a certain value, the penetration of the signal suddenly declines; that is, the signal can only be diffracted or reflected, which could make the received signal strongest. However, when the thickness continues to increase, the received signal strength will not increase but tends to be stable because the reflection and diffraction reach the extremum.

Overall, the results of these two groups of experiments show that different materials have different influences on the signal. In the experiment 1, the relationship of the three materials is plastic>cardboard>board in terms of the peak-to-peak value of the received signal. In the experiment 2, we can find that both the plastic plate and the wood plate peaked at 4.5 mm, but the paperboard plate peaked at 9 mm. According to the experimental results, these observations could be reasoned by the reflection and absorption, because different materials have different reflection and absorption characteristics for wireless signals, which could affect the channel response to a certain extent. This provides important insights into selecting substitutes for the human body.

## VI. Conclusion and Future Work

In this paper, we proposed BeSense, a device-free and real-time WiFi-based system to analyze common human behaviors (surfing, working and gaming) around computers. BeSense has been prototyped on low-cost and ubiquitous WiFi infrastructures and evaluated in extensive real-world experiments, where its performance has been verified. We envision WiFi playing a vital role in behavior recognition with its ability to offer satisfactory recognition results. This area opens many exciting and critical future research problems such as the following:

**Removing the dependence on the antenna deployment**: BeSense relies on the Fresnel zones to capture micro-gestures via manually adjusting the antenna deployment. It is quite challenging to remove such dependence to develop a system that can automatically adjust the Fresnel zones.

**Interference between gestures**: Different types of gestures interfere with each other when they alternate, and this degrades the detection performance. How to eliminate these effects is still an open issue. Moreover, whether there are other factors that affect the micro-motion experiment is still unclear.

**High-performance recognition with more behaviors**: It will be even more beneficial by detecting more daily behaviors such as drinking and eating, which have not been well addressed in the literature. The challenge lies in how to express each different behavior. Recently popular deep learning schemes may be more suitable to solve this issue and need to be studied.

**Hybrid behavior recognition**: A variety of behavior analysis schemes can be found in the literature, each of which has limitations and advantages. A hybrid system composed of multiple selected tools may result in more satisfactory results.

## Acknowledgment

This work is sponsored by the National Natural Science Foundation of China (NSFC) under Grant No. 61432004, 61772169, National Key Research and Development Program under Grant No.2018YFB0803403, the Fundamental Research Funds for the Central Universities under No.JZ2018HGPA0272.

## References

[1] K. M. Diaz et al., "Patterns of sedentary behavior and mortality in U.S. middle-aged and older adults: A national cohort study," *Ann. Internal Med.*, vol. 167, no. 7, pp. 465–475, Oct. 2017.

[2] Z. Li, Z. Feng, and J. D. Tygar, "Keyboard acoustic emanations revisited," in *Proc. ACM Conf. Computer and Communications Security (CCS)*, Alexandria, VA, Nov. 2005, pp. 373–382.

[3] S. Gupta, D. Morris, S. Patel, and D. Tan, "Soundwave: Using the doppler effect to sense gestures," in *Proc. SIGCHI Conf. Human Factors in Computing Systems*, ACM, 2012, pp. 1911–1914.

[4] J. Shotton et al., "Real-time human pose recognition in parts from single depth images," *Commun. ACM*, vol. 56, no. 1, pp. 116–124, Jan. 2013.

[5] G. Cohn, D. Morris, S. Patel, and D. Tan, "Humantenna: Using the body as an antenna for real-time whole-body interaction," in *Proc. SIGCHI Conf. Human Factors in Computing Systems*, Austin, TX, 2012, pp. 1901–1910.

[6] R. Jenke, A. Peer, and M. Buss, "Feature extraction and selection for emotion recognition from EEG," *IEEE Trans. Affective Comput.*, vol. 5, no. 3, pp. 327–339, July 2014.

[7] Y. Liu, S. Antwiboampong, J. J. Belbruno, M. A. Crane, and S. E. Tanski, "Detection of secondhand cigarette smoke via nicotine using conductive polymer films," *Nicotine Tobacco Res.*, vol. 15, no. 9, pp. 1511–1518, Mar. 2013.

[8] Y. Gu, F. Ren, and J. Li, "PAWS: Passive human activity recognition based on WiFi ambient signals," *IEEE Internet Things J.*, vol. 3, no. 5, pp. 796–805, Oct. 2016.

[9] Y. Gu, J. Zhan, Y. Ji, J. Li, F. Ren, and S. Gao, "MoSense: An RF-based motion detection system via off-the-shelf WiFi devices," *IEEE Internet Things J.*, vol. 4, no. 6, pp. 2326–2341, Dec. 2017.

[10] X. Zheng, J. Wang, L. Shangguan, Z. Zhou, and Y. Liu, "Smokey: Ubiquitous smoking detection with commercial WiFi infrastructures," in *Proc. IEEE INFOCOM 2016*, Hong Kong, Apr. 2015, pp. 17–18.

[11] Y. Wang, J. Liu, Y. Chen, M. Gruteser, J. Yang, and H. Liu, "E-eyes: Device-free location-oriented activity identification using fine-grained WiFi signatures," in *Proc. 20th Annu. Int. Conf. Mobile Computing and Networking*, Maui, HI, 2014, pp. 617–628.

[12] W. Wang, A. X. Liu, M. Shahzad, K. Ling, and S. Lu, "Understanding and modeling of WiFi signal based human activity recognition," in *Proc. ACM MobiCom*, Paris, 2015, pp. 65–76.

[13] H. Abdelnasser, K. A. Harras, and M. Youssef, "WiGest demo: A ubiquitous WiFi-based gesture recognition system," in *Proc. IEEE INFOCOM 2015*, Hong Kong, Apr. 2015, pp. 17–18.

[14] H. Li, W. Yang, J. Wang, Y. Xu, and L. Huang, "WiFinger: Talk to your smart devices with finger-grained gesture," in *Proc. ACM Int. Joint Conf. Pervasive and Ubiquitous Computing*, Heidelberg, 2016, pp. 250–261.

[15] Y. Zeng, P. H. Pathak, and P. Mohapatra, "WiWho: WiFi-based person identification in smart spaces," in *Proc. Int. Conf. Information Processing in Sensor Networks*, Vienna, 2016, p. 4.

[16] T. Xin, B. Guo, Z. Wang, M. Li, Z. Yu, and X. Zhou, "FreeSense: Indoor human identification with Wi-Fi signals," in *Proc. IEEE Global Communications Conf. (GLOBECOM)*, Washington, DC, 2016, pp. 1–7.

[17] H. Wang et al., "Human respiration detection with commodity WiFi devices: Do user location and body orientation matter?" in *Proc. ACM UbiComp*, New York, NY, Sept. 2016, pp. 25–36.

[18] J. Liu, Y. Chen, Y. Wang, X. Chen, J. Cheng, and J. Yang, "Monitoring vital signs and postures during sleep using WiFi signals," *IEEE Internet Things J.*, vol. 5, no. 3, pp. 2071–2084, June 2018.

[19] Y. Zeng, P. H. Pathak, and P. Mohapatra, "Analyzing shopper's behavior through WiFi signals," in Proc. 2nd Workshop Physical Analytics, Florence, May 2015, no. 6, pp. 13–18.

[20] S. Depatla, A. Muralidharan, and Y. Mostofi, "Occupancy estimation using only WiFi power measurements," *IEEE J. Sel. Areas Commun.*, vol. 33, no. 7, pp. 1381–1393, May 2015.

[21] C. Han, K. Wu, Y. Wang, and L. Ni, "WiFall: Device-free fall detection by wireless networks," in *Proc. IEEE INFOCOM*, Toronto, Apr. 2014, pp. 271–279.

[22] H. Wang, D. Zhang, Y. Wang, J. Ma, Y. Wang, and S. Li, "RT-Fall: A real-time and contactless fall detection system with commodity WiFi devices," *IEEE Trans. Mobile Comput.*, vol. 16, no. 2, pp. 511–526, Feb. 2017.

[23] S. Sigg, S. Shi, F. Buesching, Y. Ji, and L. Wolf, "Leveraging RF-channel fluctuation for activity recognition: Active and passive systems, continuous and RSSI-based signal features," in *Proc. Int. Conf. Advances in Mobile Computing and Multimedia*, Vienna, Dec. 2013, p. 43.

[24] H. Abdelnasser, K. A. Harras, and M. Youssef, "UbiBreathe: A ubiquitous non-invasive WiFi-based breathing estimator," in *Proc. ACM MobiHoc*, Hangzhou, June 2015, pp. 277–286.

[25] Y. Wang, K. Wu, and L. M. Ni, "WiFall: Device-free fall detection by wireless networks," *IEEE Trans. Mobile Comput.*, vol. 16, no. 2, pp. 581–594, Feb. 2017.

[26] Q. Pu, S. Gupta, S. Gollakota, and S. Patel, "Whole-home gesture recognition using wireless signals," in *Proc. ACM MOBICOM*, Miami, FL, Sept. 2013, pp. 27–38.

[27] K. Qian, C. Wu, Z. Zhou, Y. Zheng, Z. Yang, and Y. Liu, "Inferring motion direction using commodity Wi-Fi for interactive exergames," in *Proc. 2017 CHI Conf. Human Factors in Computing Systems*, Denver, CO, pp. 1961–1972.

[28] D. Wu, D. Zhang, C. Xu, Y. Wang, and H. Wang, "WiDir: Walking direction estimation using wireless signals," in *Proc. ACM Int. Joint Conf. Pervasive and Ubiquitous Computing*, Heidelberg, 2016, pp. 351–362.

[29] H. Li, K. Ota, M. Dong, and M. Guo, "Learning human activities through Wi-Fi channel state information with multiple access points," *IEEE Commun. Mag.*, vol. 56, no. 5, pp. 124–129, May 2018.

[30] J.-S. Choi, W.-H. Lee, J.-H. Lee, J.-H. Lee, and S.-C. Kim, "Deep learning based NLOS identification with commodity WLAN devices," *IEEE Trans. Veh. Technol.*, vol. 67, no. 4, pp. 3295–3303, Dec. 2018.

[31] Q. Zhou, J. Xing, W. Chen, X. Zhang, and Q. Yang, "From signal to image: Enabling fine-grained gesture recognition with commercial Wi-Fi devices," *Sensors*, vol. 18, no. 9, p. 3142, Sept. 2018.

[32] Y. Gu, X. Zhang, C. Li, F. Ren, J. Li, and Z. Liu, "Your WiFi knows how you behave: Leveraging WiFi channel data for behavior analysis," in *Proc. IEEE GLOBECOM*, Abu Dhabi, Dec. 2018, pp. 1–6.

[33] Y. Gu et al., "EmoSense: Data-driven emotion sensing via off-the-shelf WiFi devices," in *Proc. IEEE Int. Conf. Communications (ICC)*, Kansas City, MO, May 2018, pp. 1–6.

[34] J. Liu, Y. Wang, Y. Chen, J. Yang, X. Chen, and J. Cheng, "Tracking vital signs during sleep leveraging off-the-shelf WiFi," in *Proc. ACM Int. Symp. Mobile Ad Hoc Networking and Computing*, Hangzhou, 2015, pp. 267–276.